\documentclass[twocolumn]{aastex6}

\usepackage{amsmath}
\usepackage{icomma}

\def\gram{\textrm{g}}

\def\um{\mu\textrm{m}}
\def\cm{\textrm{cm}}

\def\pc{\textrm{pc}}

\def\yr{\textrm{yr}}
\def\kyr{\textrm{kyr}}
\def\Myr{\textrm{Myr}}
\def\Gyr{\textrm{Gyr}}

\def\Kelv{\textrm{K}}

\def\Lsun{\textrm{L}_{\odot}}
\def\Msun{\textrm{M}_{\odot}}
\def\MEarth{\textrm{M}_{\oplus}}

\def\mJy{\textrm{mJy}}

\def\NLife{N_{\rm life}}
\def\tanom{t_{\rm anom}}
\def\tMS{t_{\rm MS}}

\def\ga{\gtrsim}
\def\la{\lesssim}
\def\endash{\text{--}}

\shorttitle{High Rate for Boyajian's Star Anomaly}
\shortauthors{Lacki}

\begin{document}

\title{The High Rate of the Boyajian's Star Anomaly as a Phenomenon}
\author{Brian C. Lacki$^\varnothing$}
\email{astrobrianlacki@gmail.com}
\noaffiliation

\begin{abstract}
Boyajian's Star (KIC 8462852) undergoes mysterious, irregular eclipses that aren't yet explained.  It also appears to have dimmed over a time of several years, possibly decades.  I show that \emph{Kepler}'s detection of a phenomenon with a duration of $\tanom$ is only expected if it occurs at a mean rate of $\ga 30\ \Gyr^{-1} (\tanom / 100\ \yr)^{-1}$ for each \emph{Kepler} target and K2 star.  If true, the phenomenon occurs hundreds of times during the lifespan of its host stars.  Obscuration by the interstellar medium remains a plausible explanation, since it doesn't actually affect the host star.  An intervening cloud is consistent with the lack of an observed submillimeter excess but would be abnormally dilute.
\end{abstract}

\keywords{stars: individual (KIC 8462852) --- ISM: extinction --- stars: peculiar --- stars: variable: general --- extraterrestrial intelligence}

\section{Introduction}
\label{sec:Intro}
Boyajian's Star (KIC 8462852, TYC 3162-665-1) is perhaps \emph{Kepler}'s strangest discovery \citep{Boyajian16}.  Citizen scientists with the Planet Hunters project \citep{Fischer12} noticed a series of abnormal eclipses in the \emph{Kepler} photometry of this F3 dwarf \citep{Boyajian16}, which is located approximately $390\ \pc$ away \citep{Hippke16-Gaia}.  The eclipses were diverse and irregular in shape, lasted for up to several days, and blotted out up to $20\%$ of the star's flux, too long and too deep for a planetary transit.  While events with similar light curves do occur for young ``dipper'' stars with circumstellar disks, the disks usually glow with infrared radiation \citep{Cody14,Ansdell16,Scaringi16}.  Yet Boyajian's Star has no infrared excess \citep{Boyajian16,Lisse15,Marengo15}, nor any submillimeter excess \citep{Thompson16}, and its kinematics suggest that it is not newly born \citep{Boyajian16}.  

The mystery deepened when \citet{Schaefer16} claimed that the star is dimming over century timescales, using a reconstructed light curve from Harvard archival plates accessed through the Digital Access to a Sky Century @ Harvard (DASCH) project \citep{Grindlay12,Tang13}.  He found a mean flux decrease of $0.151\%\ \yr^{-1}$ from 1890 to 1989.  This claim was disputed by \citet{Hippke16-Harvard}, who noted the potential for calibration issues as the Harvard data were taken with several different telescopes and suffer the ``Menzel gap'' in the data for the 1950s and 1960s.  Further archival data at the Sonneberg Observatory appears to be consistent with a constant brightness \citep{Hippke16-Sonneberg}.  But \citet{Montet16} announced that the \emph{Kepler} light curve for Boyajian's Star showed an even faster dimming during its survey, $0.341\%\ \yr^{-1}$ for the first three years, followed by a rapid 2\% flux drop.  The star's behavior on century timescales remains unclear.

Several hypotheses have been advanced to explain Boyajian's Star, but none seem entirely satisfactory.  It is hard to accommodate both the long-term dimming and the lack of infrared and submillimeter emission \citep{Wright16-Theories}.  One hypothesis is that Boyajian's Star is experiencing a shower of debris from the breakup of a large comet, and the occulters are clusters of comet fragments and their dust tails \citep{Bodman16}.  The number of disrupted giant comets to produce the long-term dimming would be exorbitant, however \citep{Schaefer16}.  Intrinsic variability in the star is a possibility, but an atypical one for F dwarfs \citep{Wright16-Theories,Montet16}.  It is also possible that the light of Boyajian's star is being modulated by something between its system and us.  The obscuring structure's properties would be consistent with our current data on the interstellar medium (ISM) \citep{Wright16-Theories}.  An ISM scenario might also explain the apparent astrometric motion of Boyajian's star in \emph{Kepler} data, as nearby blended stars experience variable extinction \citep{Makarov16}.

The most unconventional hypothesis (so far) is that the phenomenon is artificial, the result of structures built by an alien intelligence \citep{Wright16-Aliens}.  Before \emph{Kepler} even launched, \citet{Arnold05} predicted that it might find ``megastructures'' as big as planets, which could easily signal distant societies.  On an even larger scale, it's possible that a society might shroud their entire solar system in a Dyson sphere, a swarm of satellites that collects all of the optical light of the host sun \citep{Dyson60,Kardashev64,Bradbury00}.  The long-term dimming might actually be the progress of a Dyson sphere's construction \citep{Lintott16}.  

There are multiple problems with this explanation for Boyajian's Star.  First, Dyson spheres should re-radiate the host star's emission in infrared to submillimeter wavelengths \citep{Sagan66}, but the lack of such emission from Boyajian's Star strongly constrains megastructure scenarios \citep{Wright16-Theories}.  Second, both optical and radio studies report no artificial signals from the star \citep{Abeysekara16,Schuetz16,Harp16}, despite communication being a primary motivation in \citet{Arnold05}.  

But on a more foundational level is the Fermi Paradox and its corollary, the Great Silence \citep{Brin83,Cirkovic09}.  If aliens can build megastructures that big, they are probably easily capable of launching interstellar voyages.  And \emph{Kepler} found Boyajian's Star among a few hundred thousand stars, implying that there are tens of thousands of similar systems in the Galaxy, presumably all hosting high tech societies right now, and perhaps many more over the past $4.5$ billion years. Yet the Solar System shows no evidence of artificial tampering over all that time, despite the presence of many thousands of societies of cosmic engineers.  That is the Fermi Paradox \citep{Hart75,Tipler80}.  While it's possible that actual aliens are too restrained to alter the Solar System visibly \citep{Freitas85,Scheffer94,HaqqMisra09}, the audacity of the required structures suggests that the putative Boyajian's Star inhabitants would have few such qualms.

Furthermore, all surveys for Dyson spheres have come up negative, including one examining millions of stars \citep{Slysh85,Timofeev00,Jugaku04,Carrigan09}.  More worryingly, there are no signs of galaxy-scale engineering among thousands to millions of galaxies \citep{Annis99,Griffith15,Garrett15,Zackrisson15,Lacki16}.  Even if the inhabitants of Boyajian's Star had no interest in it, the reaches of current surveys would include billions of such societies, each probably capable of galactic engineering.  Cosmic engineering seems to be an all or nothing thing, and our current surveys lean strongly towards \emph{nothing} (whether because it's infeasible or because nobody else is around).\footnote{On the subject of unconventional explanations, nobody has suggested that the phenomenon is \emph{biological} to my knowledge --- space trees of some sort.  Without knowledge of other stars and dependent on sunlight, space trees wouldn't spread beyond Boyajian's Star.  Nor would they send us narrowband radio signals or optical laser pulses. \citet{Arnold05} mentions this possibility for non-artificial, irregular transits, citing the hypothetical Kuiper Belt life of \citet{Dyson03}.}

Motivated by the strangeness of the anomaly, there have been a few attempts to find analogous stars.  \citet{LaCourse16} reports that there are no similar stars among the $165,000$ studied during the K2 campaign.  As far as long-term dimming goes, \citet{Villarroel16} find that at most $1$ star among $10^7$ appears to have disappeared from the United State Naval Observatory B1.0 catalog over the past few decades.  Meanwhile, \citet{Davenport16} advocate looking for analogs with Sloan photometry among the stars of Stripe 82.  \citet{Kochanek08} describe a survey for disappearing massive stars in nearby galaxies on decade timescales, with one apparent discovery \citep{Gerke15}, but it would only be sensitive to large secular fluctuations of the brightest stars.

This paper proceeds under the assumption that the anomalies of Boyajian's Star represent a single astronomical phenomenon.  I focus on the long-term dimming observed by \emph{Kepler} (and possibly other surveys), since it is one of the most baffling aspects of the anomaly and it has a characteristic lifetime.  I then calculate how commonly the anomaly has to occur around stars in order to have been found by \emph{Kepler}.  The anomaly turns out to be something that must happen multiple times during a typical star's lifespan.

\section{The frequency of the anomaly}
\label{sec:Rate}

\subsection{The discovery rate with \emph{Kepler}}
\label{sec:KeplerReach}
Let $\Gamma_{\rm Kepler}$ be the mean rate at which a survey like \emph{Kepler} would observe a given anomaly in one star.  It may be that only a fraction $f$ of the stars observed by \emph{Kepler} ever display the anomaly.  Then the true rate that anomalies occur around these host stars is $\Gamma_{\rm host} = \Gamma_{\rm Kepler} / f$.  Note also that $f$ does not necessarily equal the host stars' true fraction $f_{\rm true}$ among a volume-selected sample of stars.  \emph{Kepler} is biased towards Solar-type stars a few hundred parsecs away and observed Cygnus, a part of the Milky Way's disk with star-forming regions \citep{Koch10}.

The expected number of times \emph{Kepler} finds an anomalous star is 
\begin{equation}
\label{eqn:NAnom}
N_{\rm anom} = \Gamma_{\rm Kepler} N_{\rm Kepler} t_{\rm eff},
\end{equation}
where $N_{\rm Kepler}$ is the number of stars observed by \emph{Kepler}.  During its original survey, \emph{Kepler} had $150,000$ high priority target stars \citep{Batalha10}.  The extended K2 survey, studied an additional $165,000$ stars scattered over several fields along the ecliptic at a wide range of galactic latitudes \citep{Howell14,LaCourse16}.  Boyajian's Star was unique among them (expected if $N_{\rm anom} = 1$).  The effective exposure $t_{\rm eff}$ is the sum of the lifetime of the anomaly, $\tanom$, and the duration of the survey, $t_{\rm Kepler}$.  \emph{Kepler}'s original survey accumulated photometry of Boyajian's Star for $t_{\rm Kepler} = 4\ \yr$ \citep{Boyajian16}.

Thus, the frequency of the anomaly is
\begin{equation}
\label{eqn:GammaHost}
\Gamma_{\rm host} = \frac{N_{\rm anom}}{N_{\rm Kepler} t_{\rm eff} f}.
\end{equation}

\subsection{The duration of the long term decline}
\label{sec:Lifespan}
The long-term decline observed by \emph{Kepler} is probably related to the occultations.  The eclipses by themselves already appear to be a unique trait of stars in the \emph{Kepler} and K2 fields \citep{Boyajian16,Davenport16}.  But the secular trend is also quite rare, present in $\la 1\%$ of F dwarfs in the field, so the odds that it would coincidentally affect Boyajian's Star are small \citep{Lund16,Montet16}.  This suggests that both would be observed together when observing the star.

A key feature of the secular dimming is that it has a very short timescale compared to stellar evolution.  The slowest reported decline is that of \citet{Schaefer16}, in which the e-fold time is $660\ \yr$.  If the trend were continuing for much longer than that, at its current rate, then Boyajian's Star would have been implausibly bright.  It would have been a sixth magnitude star around $3,500$ years ago and visible to the naked eye, which would have been before the construction of star catalogs that deep.  Around $9,400$ years ago, Boyajian's Star would have been a million times brighter than it is now, among the brightest stars in the Galaxy.  I therefore assign a conservative upper limit of $\tanom = 10,000\ \yr$ to the anomalous dimming's lifetime.  I am assuming that the secular dimming is in fact secular, and not just a small slice of a very long period oscillation.

\emph{Kepler}'s photometry indicates an even faster dimming in recent years, with an e-folding time of $290\ \yr$.  At this rate, it would have had $V = 2.45$ around $2,500$ years ago, and its disappearance would be a glaring discrepancy with ancient star catalogs.  Yet it is unlikely that Boyajian's Star has faded anywhere near this much.  \citet{Boyajian16} are able to model the star's spectrum as a typical F dwarf at a distance of $450\ \pc$.  \emph{Gaia} has made a preliminary measurement of its distance of $390\ \pc$ \citep{Hippke16-Gaia}; the agreement between the distance modulus and the parallax distance implies that the apparent luminosity of Boyajian's Star is not greatly affected by the dimming.  The small discrepancy could be explained by a fading of $\sim 33\%$, which would take $84$ years at \emph{Kepler}'s measured rate.  The difference could be entirely due to interstellar extinction, since the measured reddening $E(B - V) = 0.11$ implies $A_V \approx 0.33$ \citep{Boyajian16}.  Thus, a medium value for the decline's lifetime is $\tanom = 100\ \yr$.

Finally, the \emph{Kepler} light curve indicates that the decline is unsteady and accelerated rapidly towards the end of the survey \citep{Montet16}.  According to \citet{Montet16}, Boyajian's Star dimmed by $2.5\%$ in a mere $200$ days.  While this precipitous fall ended afterwards, the star showed no sign of starting to recover its flux by the end of its survey.  The anomalous transits were observed over a span of about $1,500$ days during \emph{Kepler}'s survey, however, or about $4$ years \citep{Boyajian16}.  I will use a compromise value of $\tanom = 1\ \yr$ for a lower limit to the anomaly's lifespan.  The effective exposure is then $t_{\rm eff} = 5\ \yr$.

\subsection{The anomaly is common}
\label{sec:Common}
From the anomaly's lifespan and the number of stars observed during the \emph{Kepler} survey and K2, I calculate that the rate the phenomenon occurs is
\begin{multline}
\Gamma_{\rm host} = 0.32\ (0.0080 \endash 1.8)\,\Gyr^{-1}\ f^{-1}\\
\times \left(\frac{N_{\rm Kepler}}{315,000}\right)^{-1} \left(\frac{t_{\rm eff}}{10\ \kyr}\right)^{-1}
\end{multline}
from equation~\ref{eqn:GammaHost}.  The range in the parentheses is the two-sided 95\% confidence interval (one-sided 97.5\% confidence limits) \citep{Gehrels86}.  I've used the longest possible timespan allowed by Section~\ref{sec:Lifespan}.  Using a shorter duration for $t_{\rm eff}$ causes $\Gamma_{\rm host}$ to rise proportionally: it would be $32\ (0.80 \endash 180) f^{-1}\ \Gyr^{-1}$ for $t_{\rm eff} = 100\ \yr$ and $620\ (16 \endash 3,500) f^{-1}\ \Gyr^{-1}$ for $t_{\rm eff} = 5\ \yr$.

This is actually quite high when compared to the lifetime of a main sequence star.  The majority of the \emph{Kepler} target stars are dwarfs with effective temperatures near $5,500\ \Kelv$ --- they're G dwarfs like the Sun \citep{Batalha10,Ciardi11}.  Their lifetimes on the main sequence should be around $\tMS \approx 10\ \Gyr$.  Thus if every star is a potential host, it should display the phenomenon roughly $\NLife = \Gamma_{\rm host} \tMS$ times:
\begin{multline}
\NLife = 3.2\ (0.080 \endash 18)\ f^{-1}\\
\times \left(\frac{\tMS}{10\ \Gyr}\right) \left(\frac{N_{\rm Kepler}}{315,000}\right)^{-1} \left(\frac{t_{\rm eff}}{10\ \kyr}\right)^{-1}.
\end{multline}
Again, this is the most conservative possible estimate, which allows Boyajian's Star to have been a million times brighter than it is now.  A more realistic value is
\begin{multline}
\NLife = 320\ (8.0 \endash 1,800)\ f^{-1}\\
\times \left(\frac{\tMS}{10\ \Gyr}\right) \left(\frac{N_{\rm Kepler}}{315,000}\right)^{-1} \left(\frac{t_{\rm eff}}{100\ \yr}\right)^{-1}.
\end{multline}
For the shortest $t_{\rm eff}$, the fiducial $\NLife$ is $6,300\ (160 \endash 35,000)\ f^{-1}$.  It is clear that the anomaly is probably something that we would observe multiple times during a host star's lifetime, probably hundreds of times.  

If only a small fraction of stars can display the phenomenon, then in general the number of times it happens to these stars is even larger.  Boyajian's Star, with an effective temperature of $6,750\ \Kelv$ is relatively hot for a \emph{Kepler} candidate star, so perhaps the anomaly only occurs around early type stars.  Then $f$ is fairly small, although not as small as in a volume-limited population.  \citet{Batalha10} estimate that there are $24,806$ dwarf stars with effective temperatures around $6,500\ \Kelv$ among the \emph{Kepler} target stars, indicating that $f \approx 1/6$ (in fact, the great majority of these are late F dwarfs cooler than $6,500\ \Kelv$, suggesting an even smaller fraction; see \citealt{Ciardi11,Pinsonneault12}).  Against this, Boyajian's Star is expected to last shorter than a G dwarf.  With an estimated mass of $1.43\ \Msun$ and a luminosity of $4.68\ \Lsun$ \citep{Boyajian16}, its time on the main sequence is $\sim 3\ \Gyr$.  Thus,
\begin{multline}
\nonumber \NLife = \left(\frac{\tMS}{3\ \Gyr}\right) \left(\frac{f}{1/6}\right)^{-1} \left(\frac{N_{\rm Kepler}}{315,000}\right)^{-1}\\
\times \left\{ \begin{array}{l} \displaystyle 5.7\ (0.14 \endash 32)\,\left(\frac{t_{\rm eff}}{10\ \kyr}\right)^{-1}\\
                         \displaystyle 570\ (14 \endash 3,200)\,\left(\frac{t_{\rm eff}}{100\ \yr}\right)^{-1} \\
                         \displaystyle 11,000\ (290 \endash 64,000)\,\left(\frac{t_{\rm eff}}{5\ \yr}\right)^{-1}. \end{array} \right. 
\end{multline}

Restricting the host population increases $\NLife$ further.  The minimum reasonable $f$ is $\sim N_{\rm anom} / N_{\rm Kepler} = 3 \times 10^{-6}$, for which $\Gamma_{\rm host} \approx 1/t_{\rm eff}$.  This is the case when the dimming is actually a periodic variation, or an aperiodic fluctuation with a duty cycle near $1$.

\section{Interpretation}
\label{sec:Interpretations}

\subsection{Boyajian's Star and the lesson of Drake's Equation}
\label{sec:DrakeEquation}
The fundamental reason why the anomalous dimming must be a common phenomenon is familiar in the Search for Extraterrestrial Intelligence (SETI).  Equation~\ref{eqn:GammaHost} can be seen as a version of Drake's Equation, estimating the number of alien societies in the Galaxy, applied to anomalies: the number observed is the number of host stars times the fraction that are active at any given moment.  

As is well known in SETI, if a phenomenon happens only once for each star and lasts for a few decades, one would need to search millions of stars to have a reasonable chance of finding it.  Humanity has been broadcasting radio waves for that long \citep{Sullivan78} --- the same sorts of timescales found in the secular dimming of Boyajian's Star.  Since we are the only technological society we know of, and since we do not know how much longer we'll be broadcasting, we can't be sure that alien societies will last longer.  If that's the case, then there may only be a few dozen societies in the Milky Way at any moment, separated by thousands of parsecs, even if every Solar-type star gives rise to one \citep{Sagan73}.  The trouble is that we are barely capable of detecting our own radio broadcasts, at their typical levels, around the nearest stars \citep{Loeb07}.  Even using something as powerful as the Square Kilometer Array, with a potential reach of millions of stars, one would not expect to see a short-lived analog of humanity \citep{Forgan11}.  

With Boyajian's Star, we have the same problem.  \emph{Kepler} observes a measly few hundred thousand stars, not enough to catch one in a single decades-long phase (\citealt{Villarroel16} reach the same conclusion for their own anomaly search).  This means that the total amount of time a star would be observed to be anomalous must be much greater than a few decades.  Since the anomalous dimming can only be sustained for a few decades per episode, there must be several such episodes.  At least in the case of Boyajian's Star, this is allowed; for SETI, it is unlikely that a typical solar system independently generates much more than one technological society.  

SETI deals with this problem by extrapolating from humanity: it can work if alien societies last for many thousands of years, or if they produce a signal that is far more noticeable than our normal radio broadcasts, possibly something deliberately designed to grab our attention \citep{Tarter01}.  It was \emph{this very need} for an easily detectable signal that prompted \citet{Arnold05} to suggest looking for artificial transits.  In this sense, whatever is responsible for the anomaly, it is deeply linked from our point of view with the issues of SETI.  

\subsection{Implications for scenarios}
\label{sec:Implications}
If we accept the $\Gamma_{\rm host}$ calculated in Section~\ref{sec:Common}, it challenges many scenarios for Boyajian's Star.  It cannot be the result of a merger with a companion star \citep{Wright16-Theories}, since stars do not even have dozens of companions, much less devour them.  Planetary collisions might generate the required obscuration.  In our own Solar System, there were probably several large collisions (including the one responsible for the formation of the Moon) during its first $100\ \Myr$ \citep[e.g.,][]{Jackson12}.  The space motion of Boyajian's Star is unlike typical $\la 100\ \Myr$ old stars, though \citep{Boyajian16}.  Planetary collisions can happen throughout the main sequence phase of a sun if dynamical instabilities are triggered \citep{Ford01,Zhang03}.  The Nice model of the Solar System's evolution conjectures there was a chaotic phase about 4 billion years ago, for example \citep{Tsiganis05}; it may have ejected a giant planet \citep{Nesvorny11}, but in other systems collisions might be the result.  While planetary collisions are consistent with the most conservative limits on $\Gamma_{\rm host}$, the most reasonable $\Gamma_{\rm host}$ values with hundreds of events are implausibly high.  

Comet showers might happen repeatedly during a planetary system's history.  \citet{Whitmire84} and \citet{Davis84} suggested that a distant Solar companion perturbs the Oort Cloud every $26\ \Myr$, leading to a burst of comets in the inner Solar System and extinction events on Earth.\footnote{WISE later showed that the Sun has no distant stellar or brown dwarf companions \citep{Luhman14}.}  Occasional showers might also occur during encounters with passing, unbound stars \citep{Hills81}.  The moderate value of $\Gamma_{\rm host}$ associated with $\tanom = 100\ \yr$ is actually once per $30\ \Myr$, and Boyajian's Star appears to have a red dwarf companion that could act as a ``Nemesis'' \citep{Boyajian16,Bodman16}.  The comet shower hypothesis is hard to reconcile with the secular dimming trend used to derive $\tanom$, though, since roughly $0.4\ \MEarth$ of comets would have to be disrupted to fuel the obscuration for decades \citep{Schaefer16}.  There are only $40\ \MEarth$ of comets in our own Oort Cloud, suggesting the entire cloud would be depleted in showers by these events \citep{Weissman96}.  The sheer scale of the showers may be incompatible with the continued existence of complex life on Earth as well, if our Solar System is typical.  

Nor does it seem likely that we are witnessing the construction of a complete Dyson sphere.  Since repeated events are needed, a megastructure hypothesis requires that a partial shell is assembled and then disassembled, hundreds of times.  Another possibility is that the sphere is destroyed by a collisional cascade \citep{Carrigan09,Lacki16} and the inhabitants keep rebuilding it in a failure-prone state.  Moreover, this behavior needs to happen in technological societies around \emph{every} Solar-like star --- or else, it happens many thousands of times around a subset of stars.  Maybe we're in no position to question the wisdom of such powerful entities, but it strikes me as a bizarrely specific behavior for something so universal.

A further problem with many of these scenarios is that they should leave traces that last for much longer than the anomaly itself.  These traces should then be commonly found around stars.  Planetary collisions could heat the remnant worlds for thousands of years, during which they shine brightly in the infrared \citep{Zhang03}.  While the initial debris clouds from a collision of terrestrial planets might survive for only a few years \citep{Boyajian16}, a residual disk could persist for $10^3 \endash 10^7\ \yr$ afterwards as it is regenerated by collisional grinding of the fragments \citep[e.g.,][]{Jackson12}.  During this time it would be visible as an infrared excess (if not already in tension with the infrared limits for Boyajian's Star itself).  Dense, warm debris disks are rare around older stars, however.  According to \citet{Trilling08}, about $4\%$ of mature F, G, and K dwarfs ($7\%$ of mature F and A dwarfs) have $24\ \um$ excesses, although $\sim 20\%$ have $70\ \um$ excesses characteristic of exo-Kuiper Belts.  These disks reprocess $\sim 10^{-5} \endash 10^{-4}$ of the sun's light, indicating a few hundred times more dust than our asteroid belt.  A Nemesis-like comet shower is expected to last for $0.1 \endash 1\ \Myr$ \citep{Hills81,Davis84,Whitmire84}, and would have to be going on around $\sim 0.3 \endash 3\%$ of stars.  While a giant comet disintegration may be a rare episode during these showers, a shower of the required intensity could release gas and appear as time-varying absorption in the host star's spectrum \citep[e.g.,][]{Welsh13}.  For the megastructure hypothesis, one might expect a completed Dyson sphere as a result, but \citet{Carrigan09} rules out completed Dyson spheres around even $1$ Solar-like star in $10^6$.  

Two broad classes of hypotheses are unaffected by the $\Gamma_{\rm host}$ constraints.  First, it is possible that the anomalous light curve is solely due to intrinsic variability in Boyajian's Star.  The observed variability, which seems to be happening over decades, is not expected with our current understanding of stars \citep{Wright16-Theories}.  Very large amplitude dimmings lasting centuries would probably have dramatically disturbed the Earth's climate if they occurred in the Sun.  For comparison, pollution and cloud cover changes caused a few percent dimming of the amount of sunlight reaching Earth's surface happened in the 1950s to 1980s, which probably slowed down global warming in the mid 20th century \citep{Wild09}.  Still, occasional bursts of variability every $\sim 10\ \Myr$ in F dwarfs would be hard to rule out observationally without something like \emph{Kepler}.  

Second, it is possible that the anomaly has nothing to do with Boyajian's Star itself, but is due to obscuration by unassociated intervening material.  \citet{Wright16-Theories} list several possible ISM structures that could affect the star's brightness, including tiny-scale atomic structure, Bok globules, and disks around stellar remnants.  The advantage of this scenario is that it would leave no trace in the host star since nothing is actually happening there.  If this scenario is true, then no doubt the Sun displays a similar anomaly to other stars in the Galaxy without our knowing it.  \citet{Makarov16} argue that astrometric fitting of Boyajian's Star in \emph{Kepler} data also supports an ISM obscurer, because it appears to be blended with other sources that also are variable.

If the obscuration is due to the ISM, the anomaly should appear more often in stars that are distant from us and located behind large columns of gas, perhaps the molecular gas associated with star-forming regions like those in Cygnus.   A volume-limited sample of nearby stars would find disproportionately few examples of the phenomenon.   Maybe that's why K2 hasn't found anything like Boyajian's Star: it mostly observes fields away from the Galactic Plane \citep{Howell14}.  Likewise, there may be no Boyajian's Star analog among the $200,000$ stars to be examined by the Transiting Exoplanet Survey Satellite (TESS): these are nearer than the \emph{Kepler} targets and at a diverse range of galactic latitudes \citep{Ricker14}.

\subsection{Is an intervening cloud compatible with submillimeter data?}
\label{sec:Cloud}
Obscuration by a dusty cloud in the ISM is compatible with $\Gamma_{\rm host}$, and \citet{Wright16-Theories} note that small enough clouds would go unnoticed in optical light.  Dusty clouds glow at submillimeter wavelengths, however, as they are heated to $\sim 20\ \Kelv$ by background starlight \citep{Draine11}.  One test of the intervening cloud test is submillimeter observations.

The thermal flux at a frequency $\nu$ from a cloud with a temperature of $T$ is
\begin{equation}
F_{\nu} = \frac{2 h \nu^3 \Omega}{c^2} \frac{1 - e^{-\tau_{\nu}}}{e^{-h \nu/(k_B T)} - 1} \approx \frac{2 h \nu^3 \Omega}{c^2} \frac{\tau_{\nu}}{e^{-h \nu/(k_B T)} - 1},
\end{equation}
for a cloud covering a solid angle $\Omega$ of the sky with an absorption optical depth at $\nu$ of $\tau_{\nu}$.  The other constants are $c$, the speed of light; $h$, Planck's constant; and $k_B$, Boltzmann's constant.

We can estimate the minimum size of the cloud from the proper motion of Boyajian's Star, $\mu = 0.0153 \arcsec\ \yr^{-1}$ \citep{Zacharias13}, as long as the motions of the cloud and Boyajian's Star are uncorrelated.  Then the angular width of the cloud is $\theta \ga \mu \tanom$, and the solid angle covered is $\Omega \approx \theta^2 \ga (\mu \tanom)^2$.  Assuming $\tanom \approx 100\ \yr$, $\theta \ga 1.5 \arcsec$, and $\Omega \ga 2.3\ \sq \arcsec$.  A minimal cloud would be unresolved by the Submillimeter Array, which observed the star with a beam size of $4\arcsec \times 3\arcsec$ \citep{Thompson16}.  The cloud must be smaller than $\sim 1 \arcmin$ in diameter, since there are non-anomalous stars at that angular distance \citep{Wright16-Theories}.

The optical depth of the cloud at visible wavelengths is approximately the observed fractional dimming of the star's brightness.  To extrapolate this optical depth to submillimeter wavelengths, I use the extinction curves given in \citet{Fitzpatrick07}.  At an infrared wavelength $\lambda$, these curves can be described with a function
\begin{equation}
k(\lambda - V) = (0.63 R_V - 0.83) \left(\frac{\lambda}{\um}\right)^{-1.84} - R_V,
\end{equation}
using a reddening coefficient of $R_V = 3.001$ \citep{Fitzpatrick07}.  The ratio of $\tau_{\nu}$ to $\tau_V$ in $V$-band is then
\begin{equation}
\frac{\tau_{\nu}}{\tau_V} = \frac{1 - \omega_{\nu}}{1 - \omega_V} \frac{k(\lambda - V) + R_V}{R_V},
\end{equation}
where $\omega_V = 0.5$ is the albedo of interstellar dust in V-band and $\omega_{\nu} = 0$ is the albedo at submillimeter wavelengths \citep{Draine11}.  For reference, I also consider the case of black dust with equal opacity at all frequencies.

\begin{deluxetable}{lllll}
\tablecolumns{3}
\tablecaption{Expected submillimeter flux from a minimal obscuring cloud\label{table:CloudSubmm}}
\tablehead{\colhead{$\lambda$} & \colhead{$\tau_{\nu}/\tau_V$} & \colhead{$F_{\nu}$} & \colhead{$F_{\nu}^{\rm black}$} & \colhead{$F_{\rm max}$}\\ \colhead{($\um$)} & & \colhead{($\mJy$)} & \colhead{($\mJy$)} & \colhead{($\mJy$)}}
\startdata
$1100$ & $1.8 \times 10^{-6}$ & $6.2 \times 10^{-5}$ & $35$ & $32.1$\\
$850$  & $2.9 \times 10^{-6}$ & $1.5 \times 10^{-4}$ & $51$ & $2.55$\\
$450$  & $9.3 \times 10^{-6}$ & $7.1 \times 10^{-4}$ & $77$ & $2.19$
\enddata  
\tablecomments{I assume $T = 20\ \Kelv$, $\Omega = (1.5\arcsec)^2$, and $\tau_V = 0.03$.  $F_{\nu}$ is the predicted submillimeter flux using the \citet{Fitzpatrick07} extinction law.  $F_{\nu}^{\rm black}$ is the submillimeter flux if $\tau_{\nu} = \tau_V$.  The observed flux limits from \citet{Thompson16} are listed under $F_{\rm max}$.}
\end{deluxetable}

As seen in Table~\ref{table:CloudSubmm}, the predicted submillimeter fluxes of a $20\ \Kelv$ cloud with $\tau_V = 0.03$ (as required for the \emph{Kepler} dimming) is easily consistent with the submillimeter flux limits \citep{Thompson16}.  If the cloud has the minimum required size and density, it is $\sim 3,000$ times too faint to detect.  If I use $\tanom = 10,000\ \yr$ to calculate $\theta$, then the cloud is about $3$ times the flux limits.  An unphysical black cloud should have been detected by the Submillimeter Array.

If molecular, the mass of a minimal cloud would be abnormally small compared to known structures.  An extinction-to-gas ratio of $A_V/N_H = 5.3 \times 10^{-22}\ \cm^2\ \textrm{H}^{-1}$ implies a gas mass of $2 \times 10^{27}\ \gram \approx 0.3\ \MEarth$ for $A_V \approx 0.03$, assuming a distance of $200\ \pc$ \citep{Draine11}.  A cloud that dilute could not be gravitationally bound.  The maximum mass allowed by the submillimeter data for an unresolved cloud is $\sim 6 \times 10^{30}\ \gram \approx 0.003\ \Msun$.  

Transient clumps of gas are generated in molecular clouds by supersonic turbulence \citep{MacLow04}, but most such clumps follow the \citet{Larson81} relation between velocity dispersion and length scale.  With a diameter of $\sim 0.0015\ \pc$, an extrapolation of Larson's relations for a self-gravitating clump gives a mass of $\sim 0.002\ \Msun$, requiring the cloud to be almost detectable \citep{Draine11}.  Put another way, the characteristic surface density of Galactic molecular clumps is $\sim 200\ \Msun\ \pc^{-2}$ \citep{Solomon87}, whereas a minimal cloud has $0.5\ \Msun\ \pc^{-2}$.  A typical density cloud is allowed by the submillimeter data, but the mass must be hidden in sub-clumps to avoid obscuring Boyajian's Star too much.  Perhaps sub-clumps are responsible for the larger eclipses, although they would have to be extremely small and themselves would be dilute.  In a Kolmogorov turbulence cascade, density contrasts are expected to be negligible on small scales (\citealt{Elmegreen04}; this is also the case for plasma turbulence, as in \citealt{Armstrong95}).  Supersonic turbulence can generate sharp features like shocks, though \citep{Elmegreen04}.

An intervening disk explanation may better explain the size of the cloud \citep{Wright16-Theories}.

\subsection{Alternatives to a common anomaly}
\label{sec:Caveats}
There are a couple of caveats to my calculation of $\Gamma_{\rm host}$, and its favoring of ISM obscuration.

The main loophole is if $\tanom$ is actually much longer than $10,000\ \yr$.  In order to be true, the apparently secular dimming must actually be an oscillation on century (or longer) timescales, which only appears to be monotonic because we only have data over a century.  I assumed that the dimming trend is non-repeating because the light curve appears to be roughly constant for the early 20th century \citep[as in][]{Hippke16-Sonneberg}.  Most of the dimming is concentrated in the last few years, as observed with \emph{Kepler} \citep{Montet16}.  But a long period of constant flux punctuated by drops would be the expected light curve if Boyajian's Star is eclipsed by a distant companion's disk (although an improbable case according to \citealt{Wright16-Theories}).  If it is shown that Boyajian's Star was brightening decades ago, this would be evidence that the dimming is actually an oscillation.  Then the relevant timescale would be the duration the star is variable, which could be millions of years.

The anomaly might be observed as a fluke even if the actual value of $\Gamma_{\rm host}$ is much smaller than calculated.  If the true value is $p \ll 1$ times the derived value, then the probability that the anomaly would be observed in the \emph{Kepler} field is $\sim p$.  For example, the anomaly is consistent with being a singular event in main sequence stars ($\NLife = 1$) at $p = 1/300$.  This is a $3\ \sigma$ deviation --- very unlikely, but not unheard of.

The other possibility for a fluke is that the secular dimming is unrelated to the eclipses of Boyajian's Star.  Then, the derived $\Gamma_{\rm host}$ would still apply to the dimming but not to the eclipses, which could have been happening for millions of years.  Of $644$ F dwarfs with good photometry in the DASCH Harvard data, \citet{Lund16} found that $11$ of them ($1.7\%$) of them have significant variability on century timescales.  Boyajian's Star is not one of these $11$, so the number of stars with enough anomalous variability may be greater.  \citet{Montet16} found that the level of variability in Boyajian's Star over the \emph{Kepler} mission is very rare.  The rapid trend in the brightness for the first three years is found in $0.7\%$ of F stars, and the sharp fall-off of the last year is unique in the several hundred F stars they examine \citep{Montet16}.  Again, it is unlikely but not unheard of that the dimming is an effect expected $0.1 \endash 1\%$ of the time and is unrelated to the unexplained eclipses.  More information about the variability of F dwarfs on long timescales would be helpful in this regard.

\acknowledgments
I wish to acknowledge the use of NASA's Astrophysics Data System and arXiv.  I also used the VizieR catalogue access tool, CDS, Strasbourg, France to get the proper motion of Boyajian's Star.

\end{document}